\documentclass[conference]{IEEEtran}%
\pdfoutput=1
\usepackage{graphicx,multicol} 
\usepackage{epsfig} 
\usepackage{amsmath} 
\usepackage{amssymb}  
\usepackage{amsthm}
\usepackage{mathtools}
\usepackage{float}
\usepackage{mathabx}
\usepackage{bm}
\usepackage{enumitem}

\newtheorem{theorem}{Theorem}

\newcommand{\ONECOLUMN}[1]{}
\usepackage{cases,setspace,adjustbox,xspace}

\usepackage{texdef2015}

\usepackage{cite}

\ifCLASSINFOpdf
\else
\fi
\usepackage{algorithmic}
\usepackage[ruled,vlined]{algorithm2e}

\usepackage{array}
\usepackage{makecell}
\newcolumntype{L}{r>{\centering\arraybackslash}m{.5cm}}

\usepackage{amstext} 

\usepackage[update,prepend]{epstopdf}

\usepackage{eqparbox}

\usepackage[tight,footnotesize]{subfigure}

\usepackage{fixltx2e}
\usepackage{url}

\newcommand{\age}{\Delta}
\newcommand{\Lcal}{\mathcal{L}}
\newcommand{\Qcal}{\mathcal{Q}}

\newcommand{\R}{\mathbb{R}}
\newcommand{\dq}[2][q]{\delta_{#2,#1}}
\newcommand{\dqt}[2][q(t)]{\delta_{#2,#1}}
\newcommand{\ql}{q_l}

\newcommand{\laml}[1][l]{\lambda^{(#1)}}

\newcommand{\cvec}[1]{\Bracket{#1}}

\newcommand{\smvec}[1]{\left[\begin{smallmatrix} #1\end{smallmatrix}\right]}
\newcommand{\rowvec}[1]{\left[\begin{matrix} #1\end{matrix}\right]}

\newcommand{\smrvec}[1]{\setlength\arraycolsep{2pt}\rowvec{#1}}
\newcommand{\rvec}[1]{\smrvec{#1}}

\newcommand{\lambdaOne}{\lambda_i}
\newcommand{\lambdaTwo}{\lambda_{-i}}
\newcommand{\n}{m}

\newcommand{\piv}{\text{\boldmath{$\pi$}}}

\newcommand{\pibar}{\bar{\pi}}
\newcommand{\vvbar}{\bar{\vv}}

\newcommand{\zerov}[1][\mbox{}]{\mathbf{0}_{#1}}
\newcommand{\onev}[1][n]{{\mathbf{1}}_{#1}}

\newcommand{\coeffOne}{c}

\newcommand{\barv}{\bar{v}}

\newcommand{\resetMapArrSelf}[1]{\Lambda_{#1}}
\newcommand{\resetMapArrOther}[1]{\Gamma_{#1}}
\newcommand{\resetMapArrDepart}[1]{\mathbf{D}_{#1}}

\renewcommand{\vec}[1]{\begin{bmatrix} #1\end{bmatrix}}
\renewcommand{\bv}{\mathbf{b}}

\usepackage{tikz}
\usetikzlibrary{automata,arrows,positioning,calc,shapes.multipart,chains,shapes,decorations}

\begin{document}
%
\title{Timely Updates By Multiple Sources: The M/M/1 Queue Revisited}
\author{\IEEEauthorblockN{Sanjit~K.~Kaul}
\IEEEauthorblockA{Wireless Systems Lab, IIIT-Delhi \\
email: skkaul@iiitd.ac.in}
\and
\IEEEauthorblockN{Roy~D.~Yates}
\IEEEauthorblockA{WINLAB, ECE Dept., Rutgers University\\
email: ryates@winlab.rutgers.edu}}
\IEEEoverridecommandlockouts
\maketitle

\begin{abstract}
Multiple sources submit updates to a monitor  through an M/M/1 queue.  A stochastic hybrid system (SHS) approach is used to derive the average age of information (AoI)  for an individual source as a function of the offered load  of that source and the competing update traffic offered by other sources. This work corrects an error in a prior analysis.  By numerical evaluation, this error is observed to be small and qualitatively insignificant.
\end{abstract}
%

\IEEEpeerreviewmaketitle
\section{Introduction}
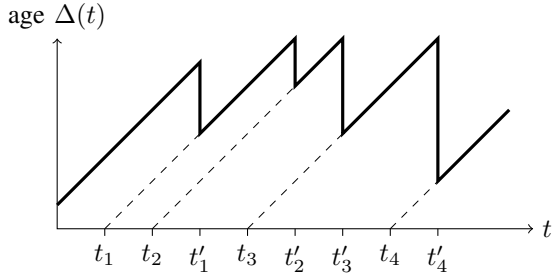
\begin{figure}
	\begin{center}
	\begin{tikzpicture}[baseline=(current bounding box.center),scale=\linewidth/28cm]
\draw [<->] (0,8) node [above] {age $\age(t)$} -- (0,0) -- (20,0) node [right] {$t$};
\draw [very thick] (0,1) -- (6,7) -- (6,4)  -- (10,8) 
-- (10,6)  -- (12,8) -- (12,4) -- (16,8) -- (16,2) -- (19,5);
\draw [ultra thin, dashed] (2,0) to (6,4)  (4,0) to (10,6) (8,0) to (12,4) (14,0) to (16,2);
\draw 
(2,0) -- (2,-0.3) node [below] {$t_1$} 
(4,0) --(4,-0.3) node [below] {$t_2$} 
(6,0) -- (6,-0.3) node [below] {$t'_1$}
(8,0) --(8,-0.3) node [below] {$t_3$}
(10,0) --(10,-0.3) node [below] {$t'_2$}
(12,0) -- (12,-0.3) node [below] {$t'_3$}
(14,0)--(14,-0.3)  node [below] {$t_4$}
(16,0) -- (16,-0.3) node [below] {$t'_4$};
\end{tikzpicture}
	\caption{Updates arrive at  times $t_k$ and are received by the monitor at times $t'_k$. At instants $t'_k$, the age is reset to $t'_k - t_k$, which is the time spent by the update in the system. Age increases linearly otherwise.}
	\label{fig:age}	
	\end{center}
\vspace{-5mm}
\end{figure}

Fueled by ubiquitous connectivity and advancements in portable devices, \emph{real-time status} updates have become increasingly popular. 
This has driven emerging analytical interest in the Age of Information (AoI) metric for characterizing the timeliness of these updates \cite{Kaul-YG-infocom2012}.  Specifically, an update packet with time-stamp $u$ is said to have age $t-u$ at a time $t\ge u$.  When the monitor's freshest received update at time $t$ has time-stamp $u(t)$,  the  age is the random sawtooth process $\age(t)=t-u(t)$ depicted in Figure~\ref{fig:age}. Optimization based on AoI metrics of both the network and the senders' updating policies  has yielded new and even surprising results \cite{Sun-UBYKS-IT2017UpdateorWait,Yates-isit2015}.

 Prior age analyses of the M/M/1 queue are based on graphical analysis of the sawtooth age process. As indicated by the title, this work revisits age analysis for updates from multiple sources passing through the M/M/1 queue using the stochastic hybrid systems (SHS) method. Prior use of SHS for age analysis has been restricted to systems in which old updates in the system are purged when new updates arrive. This resulted in a finite discrete state space, a technical requirement  of SHS analysis. The challenge in this work is to apply SHS to an M/M/1 queue in which the discrete state, namely the queue backlog, can  be arbitrarily large. Our approach is to employ SHS  for age analysis of the M/M/1/m queue in which the system discards arriving updates that find $m$ previously queued updates and then allow $m\to\infty$.  
 
 The main result, \Thmref{ageSourcei}, can be shown to be numerically identical to the independently derived result in \cite[Equation (55)]{moltafet2019age}.  In fact, both  \Thmref{ageSourcei} and \cite[Equation (55)]{moltafet2019age} correct a faulty age analysis of the multi-source M/M/1 queue in \cite{Yates-Kaul-isit2012} that propagated to \cite{Yates-Kaul-IT2018}.
\subsection{Prior Work}
\label{sec:related}
AoI analysis of updating systems started with the analyses of status age in single-source single-server queues \cite{Kaul-YG-infocom2012}, the M/M/1 LCFS queue with preemption in service \cite{Kaul-YG-ciss2012}, and the M/M/1 FCFS system with multiple sources \cite{Yates-Kaul-isit2012}.  
Since these initial efforts, there have been a large number of  contributions to AoI analysis. This section summarizes work related to the analysis of age, with an emphasis on systems in which updates arrive as stochastic processes to queues and networks.
 
 To evaluate AoI for a single source sending updates through a network cloud \cite{Kam-KE-isit2013random} or through an M/M/2 server \cite{Kam-KE-isit2014diversity,Kam-KNE-IT2016diversity}, out-of-order packet delivery was the key analytical challenge. A related (and generally more tractable) metric, peak age of information (PAoI), was introduced in \cite{Costa-CE-isit2014}. 
Properties of PAoI were also studied for various M/M/1 queues that support preemption of updates in service or discarding of updates that find the server busy \cite{Costa-CE-IT2016management,Kavitha-AS-arxiv2018}.  In  \cite{Costa-CE-isit2014,Costa-CE-IT2016management}, the authors analyzed AoI and PAoI for 
queues that discard arriving updates if the system is full and also for  a third queue
in which an arriving update would preempt a waiting update.  

For a single updating source, distributional properties of the age process were analyzed for the D/G/1 queue under FCFS \cite{Champati-AG-aoi2018}, as well as for single server FCFS and LCFS queues  \cite{Inoue-MTT-arxiv2017}.  Packet deadlines were   found to improve AoI \cite{Kam-KNWE-isit2016deadline}. Age-optimal preemption policies were identified for updates with deterministic service times \cite{Wang-FY-arxiv2018}.  AoI was evaluated in the presence of packet erasures at the M/M/1 queue output \cite{Chen-Huang-isit2016} and for memoryless arrivals to a two-state Markov-modulated service process \cite{Huang-Qian-globecom2017}. 

There have also been efforts to evaluate and optimize age for multiple sources sharing a queue or simple network \cite{Huang-Modiano-isit2015,Kadota-UBSM-Allerton2016,Kaul-Yates-isit2017,Yates-Kaul-IT2018, Najm-Telatar-aoi2018,Jiang-KZZN-isit2018,moltafet2020average}. In \cite{Yates-Kaul-IT2018}, the SHS approach was introduced to extend AoI results to preemptive queues with multiple sources.  SHS has also been employed in \cite{moltafet2020average} to evaluate a two-source M/M/1 queue in which the most recent update from each source is queued.  With synchronized arrivals, a maximum age first policy was shown to be optimal under preemption in service and near-optimal for non-preemptive service \cite{Sun-UBK-aoi2018}. A similar maximum age matching approach  was analyzed for an orthogonal channel system \cite{Tripathi-Moharir-globecom2017}. Scheduling based on the Whittle index was also shown to perform nearly optimally \cite{Kadota-UBSM-Allerton2016,Jiang-KZN-itc2018,Hsu-isit2018}.  AoI analysis of preemptive priority service systems has also been explored \cite{Najm-NT-arxiv2018,Kaul-Yates-isit2018priority}.
Updates through  communication channels have also been studied, including  hybrid ARQ  \cite{Parag-TC-wcnc2017realtime,Najm-YS-isit2017,Yates-NSZ-isit2017,Ceran-GG-wcnc2018,Sac-BUBD-spawc2018} for channels with bit erasures, and channel coding by an energy harvesting source \cite{Baknina-Ulukus-arxiv2018coded}.

%
The first evaluation of the average AoI over multihop network routes \cite{Talak-KM-allerton2017} employed a discrete-time version of the  status sampling network subsequently introduced in \cite{Yates-arxiv2018networks}.
When multiple sources employ wireless networks subject to interference constraints, AoI has been analyzed under a variety of link scheduling strategies \cite{He-YE-IT2018,Lu-JL-mobicom2018,Talak-KM-arxiv2018distributed,Talak-KM-arxiv2018perfectCSI,Talak-KKM-arxiv2018}.
When update transmission times 
are exponentially distributed,  sample path arguments have shown that a preemptive Last-Generated, First-Served (LGFS) policy results in smaller age processes at all nodes of the network than any other causal policy~\cite{Bedewy-SS-isit2016,Bedewy-SS-isit2017,Bedewy-SS-arxiv2017}. 
\subsection{System Model}

\begin{figure}[t]
	\centering
	\begin{tikzpicture}[node distance=0.95cm]
	\node (vdots) {\raisebox{1.1pt}{$\vdots$}};
	\node[draw,circle,thick] (s1)[above of = vdots] {\small\shortstack{Source\\ 1}};
	\node[draw,circle,thick] (s2)[below of=vdots] {\small \shortstack{Source\\ $N$}};
	\node (emid)[right of=vdots] {};
	\node (enter)[right of=emid] {};
	\node (smid)[right of=enter] {};
	\draw [->, thick]  (s1.east) --  node [above right] {$\lambda_1$} (enter);
	\draw [->, thick]  (s2.east) -- node [below right] {$\lambda_N$}(enter);
	\node[draw,circle,thick] (server)[right of=smid] {$\mu$};
	\draw[thick] (server.west) -- ++(0,0.5cm) --  ++(-1.75cm,0);
	\draw[thick] (server.west) -- ++(0,-0.5cm) --  ++(-1.75cm,0);
	\foreach \i in {1,...,3}
	\draw[thick] (server.west) ++(-\i*0.5cm,0.5cm) -- ++(0,-1cm);
	\node (monmid)[right of=server] {};
	\node[draw,circle,thick] (mon)[right of=monmid] {\small Monitor};
	\draw [->,thick] (server.east) -- (mon.west);
	\end{tikzpicture}
	\caption{$N$ sources send updates through a shared queue to a monitor.}
	\label{fig:twosourcespic} 
	\vspace{-3mm}
\end{figure}
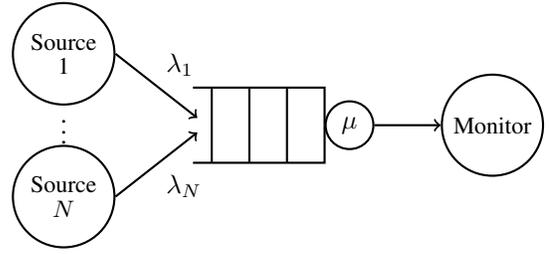

Status updates from $N$ independent sources are queued in a first-come-first-served (FCFS) manner at a single server facility.  On completion of service, an update  is delivered to a monitor, as illustrated in  Figure~\ref{fig:twosourcespic}.  Updates of source $i$ arrive as a Poisson process of rate $\lambda_i$ and service times are exponentially distributed with rate $\mu$.  The load offered by user $i$ is $\rho_i = \lambda_i/\mu$.  The total offered load is $\rho = \sum_{i=1}^{N} \rho_i$ %
and the corresponding total rate of arrivals of updates is $\lambda$. For queue stability we require $\rho < 1$. 

Each update is associated with a timestamp that records the time it was generated by its source. For this work, it is also the time the update arrives at the facility. At time $t$, let $u_i(t)$ be the timestamp of an update of source $i$ most recently received by the monitor. The age of updates from source $i$ at the monitor is the stochastic process $\age_i(t) = t - u_i(t)$. Figure~\ref{fig:age} shows an illustration. We will calculate the average age $\age_i = \lim_{t\to\infty} \E{\age_i(t)}$. All sources other than $i$ together constitute a Poisson process of rate $\lambda_{-i} = \sum_{j\ne i} \lambda_j$ and a corresponding offered load of $\rho_{-i} = \lambda_{-i}/\mu = \rho - \rho_i$.

\section{Stochastic Hybrid Systems for AoI}
\label{sec:shs-intro}
As in \cite{Yates-Kaul-IT2018}, we model the system as a stochastic hybrid system (SHS) with hybrid state $(q(t),\xv(t)$). Here,
$q(t)$ is discrete and represents a Markov state sufficient for the characterization of age, while the row vector $\xv(t)\in\R^{n+1}$ is continuous and captures the evolution of $n+1$ age-related processes. 

In this work, we use the SHS described in \cite{Hespanha-2006modelling} in a simplified form in which $\xv(t)$ is a piecewise linear process. We now summarize this simplified SHS; further details can be found in \cite{Yates-Kaul-IT2018} and references therein. 

The discrete state $q(t)\in \Qcal=\{0,\ldots,m\}$ is the number of update packets in the system. 
In the graphical representation of the Markov chain $q(t)$, each state $q\in\Qcal$ is a node and each transition $l$ is a directed edge $(q_l,q'_l)$ with transition rate $\laml\dqt{q_l}$. Note that the Kronecker delta function $\dq{q_l}$ ensures that transition $l$ occurs only in state $q_l$. For each transition $l$, there is a transition reset mapping that can induce discontinuous jumps in the continuous state $\xv(t)$. For AoI analysis, we employ a linear mapping 
of the form $\xv'=\xv\Amat_l$.  That is, transition $l$ causes the system to jump to discrete state $q'_l$ and resets the continuous state from $\xv$ to $\xv'=\xv\Amat_l$. For tracking of the age process, the transition reset maps are binary: $\Amat_l\in\set{0,1}^{(n+1)\times (n+1)}$. 
Moreover, in each discrete state $q(t) = q$, the continuous state evolves as
\begin{align}\eqnlabel{xv-evolution}
\dot{\xv}(t)=\bv_q.
\end{align}

In using a piecewise linear SHS for AoI, the elements of $\bv_q$ will be binary. We will see that the ones in $\bv_q$ correspond to certain relevant components of $\xv(t)$ that grow at unit rate in state $q$ while the zeros mark components of $\xv(t)$ that are irrelevant in state $q$ to the age process and need not be tracked.

The transition rates $\set{\laml}$ correspond to the transition rates associated with the continuous-time Markov chain for the discrete state $q(t)$; but there are some differences. Unlike an ordinary continuous-time Markov chain, the SHS may include self-transitions in which the discrete state is unchanged because a reset occurs in the continuous state. Furthermore, for a given pair of states $i,j\in\Qcal$, there may be multiple transitions $l$ and $l'$ in which the discrete state jumps from $i$ to $j$ but the transition maps $\Amat_l$ and $\Amat_{l'}$ are different.

It will be sufficient for average age analysis 
to define for all $\qhat\in\Qcal$,
\begin{subequations}\eqnlabel{pivi-defns}
\begin{align}
\pi_{\qhat}(t) &= 
\E{\dqt{\qhat}},
\eqnlabel{piqhat-defn}\\
v_{\qhat j}(t)&=
\E{x_j(t)\dqt{\qhat}},\quad  0 \le j \le n,\eqnlabel{vqi-defn}\\
\shortintertext{and the vector functions}
 \vv_{\qhat}(t)&=\cvec{v_{\qhat 0}(t),\dots,v_{\qhat n}(t)}=\E{\xv(t)\dqt{\qhat}}. \eqnlabel{vv-defn}
 \end{align}
\end{subequations}

Note that $\pi_{\qhat}(t)=\E{\dqt{\qhat}}=\prob{q(t)=\qhat}$ denote the discrete state probabilities.
Similarly, $\vv_{\qhat}(t)$ is the conditional expectation of the age process, given that $q(t) = \qhat$, weighed by the probability of being in $\qhat$.

Let $\Lcal$ be the set of all transitions. For each state $q$, let
\begin{equation}
\Lcal'_{q}=\set{l\in\Lcal: q'_l=q}\quad\text{and}\quad \Lcal_{q}=\set{l\in\Lcal: q_l=q}
\end{equation}
denote the respective sets of incoming and outgoing transitions. A foundational assumption for age analysis is that the Markov chain $q(t)$ is ergodic; otherwise, time-average age analysis makes little sense. Under this assumption, the state probability vector $\piv(t)=\rvec{\pi_0(t) &\cdots&\pi_m(t)}$ always  converges to the unique stationary vector $\bar{\piv}=\rvec{\pibar_0&\cdots&\pibar_m}$ satisfying
\begin{subequations}
\begin{align}
\bar{\pi}_{\qbar}\sum_{l\in\Lcal_{\qbar}}\laml&=\sum_{l\in\Lcal'_{\qbar}}\laml\bar{\pi}_{\ql},\quad \qbar\in\Qcal,\eqnlabel{AOI-SHS-pi}\\
\sum_{\qbar\in\Qcal}\bar{\pi}_\qbar&=1.\eqnlabel{AOI-SHS-pi-sum}
\end{align}
\eqnlabel{AOI-SHS-pi-all}
\end{subequations}
When $\piv(t)=\bar{\piv}$,    $\vv(t)=\rvec{\vv_0(t)&\cdots&\vv_m(t)}$  has been  shown \cite{Yates-Kaul-IT2018}  to obey a system of first order differential equations, such that for all $\qbar\in\Qcal$,
\begin{align}
\dot{\vv}_{\qbar}(t)&=
\bv_{\qbar}\pibar_{\qbar}+\sum_{l\in\Lcal'_{\qbar}}\laml \vv_{\ql}(t)\Amat_l
-\vv_{\qbar}(t)\sum_{l\in\Lcal_{\qbar}}\laml.
\eqnlabel{vv-derivs-pibar}
\end{align}
%
Depending on the  reset maps $\Amat_l$, the differential equation \eqnref{vv-derivs-pibar} may or may not be stable. However,  when \eqnref{vv-derivs-pibar} is stable, 
each $\vv_{\qbar}(t)=\E{\xv(t)\dqt{\qbar}}$ converges to a limit $\vvbar_{\qbar}$ as $t\goes\infty$. 
In this case, it  follows 
that
\begin{align}
\E{\xv}&=\limty{t}\E{\xv(t)}
=\limty{t}\sum_{\qbar\in\Qcal} \E{\xv(t)\delta_{\qbar,q(t)}}
=\sum_{\qbar\in\Qcal} \bar{\vv}_{\qbar}.\nonumber
\end{align}
Here we follow the convention adopted in \cite{Yates-Kaul-IT2018} that $x_0(t)=\age(t)$ is the age at the monitor. With this convention, the average age of the process of interest is then
$\age=\E{x_0}=\sum_{\qbar\in\Qcal}\vbar_{\qbar 0}$.
The following theorem provides a simple way to calculate the average age in an ergodic queueing system.
\begin{theorem}\thmlabel{AOI-SHS}
\cite[Theorem~4]{Yates-Kaul-IT2018}
If the discrete-state Markov chain $q(t)$ is ergodic with stationary distribution $\bar{\piv}$ and we can find a non-negative solution $\vvbar=\rvec{\vvbar_0&\cdots\vvbar_m}$ 
such that 
\begin{align}
\bar{\vv}_{\qbar}\sum_{l\in\Lcal_{\qbar}}\laml &=\bv_{\qbar}\bar{\pi}_{\qbar}+ \sum_{l\in\Lcal'_{\qbar}}\laml \bar{\vv}_{\ql}\Amat_l,\qquad \qbar\in\Qcal,\eqnlabel{AOI-SHS-v}
\end{align}
then
the differential equation \eqnref{vv-derivs-pibar} is stable and 
the average age of the AoI SHS is given by $\age=\sum_{\qbar\in\Qcal} \vbar_{\qbar 0}$.
\end{theorem}


\section{SHS Analysis of Age}
\begin{figure}[t]
	\centering
	\begin{tikzpicture}[->, >=stealth', auto, semithick, node distance=1.8cm]
	\tikzstyle{every state}=[ellipse,fill=white,draw=black,thick,text=black,scale=1]
	\tikzstyle outflow=[semithick,draw=black,dotted]
	\node[state]    (0)   {$0$};	
	\node[state]    (1)[right of=0]   {$1$};
	\node[state]    (2)[right of=1]   {$2$};	
	\node[circle]   (3)[right of=2]   {$\ldots$};	
	\node[state]    (m)[right of=3]   {$\n$};
	\path
	
	(0) edge[above, bend left= 30]     node{$\lambdaOne$}     	(1)	
	(0) edge[above, bend left= 80]     node{$\lambdaTwo$}     	(1)	
	(1) edge[below, bend left= 30]     node{$\mu$}     	(0)
	(1) edge[above, bend left= 30]     node{$\lambdaOne$}     	(2)	
	(1) edge[above, bend left= 80]     node{$\lambdaTwo$}     	(2)		
	(2) edge[below, bend left= 30]     node{$\mu$}     	(1)
	(2) edge[above, bend left= 30]     node{$\lambdaOne$}     	(3)	
	(2) edge[above, bend left= 80]     node{$\lambdaTwo$}     	(3)		
	(3) edge[below, bend left= 30]     node{$\mu$}     	(2)
	(3) edge[above, bend left= 30]     node{$\lambdaOne$}     	(m)	
	(3) edge[above, bend left= 80]     node{$\lambdaTwo$}     	(m)		
	(m) edge[below, bend left= 30]     node{$\mu$}     	(3);						
	\end{tikzpicture}
	\caption{The SHS Markov chain for transitions between discrete states.}
	\label{fig:CTMC}
	\vspace{-5mm}
\end{figure}
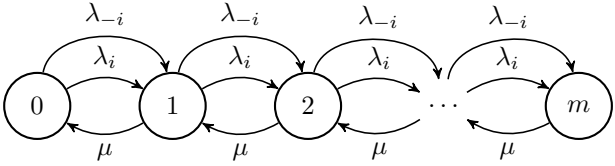

The discrete Markov state $q(t)\in \{0,1,2,\ldots\}$ tracks the number of updates in the system at time $t$. Figure~\ref{fig:CTMC} shows the evolution of the state for a corresponding blocking system that has a finite occupancy of $\n$. In states $0 < k < \n$, an arrival, which results from a rate $\lambda$ transition, increases the state to $k+1$. A departure on completion of service, which results from a rate $\mu$ transition, reduces the occupancy to $k-1$. An empty system, $q(t) = 0$, can only see an arrival; while when the system is full, $q(t) = \n$, all arrivals are blocked and cleared and only a departure may take place.

Let $\age_i^{\n}$ be the average age of source $i$ for the size $m$ blocking system. The average age $\age_i$ of source $i$ at the monitor can be obtained as
\begin{align}
\age_i = \lim_{\n\to \infty} \age_i^{\n}.
\eqnlabel{age_as_limiting}
\end{align}
For the blocking system,  $\xv(t)=\rvec{x_0(t)& x_1(t) & \ldots & x_\n(t)}$ is the age state. Here $x_0(t)$ is the age process of source $i$ at the monitor. The age process $x_j(t)$, $1 \le j \le k$, tracks the age to which the age $x_0(t)$ will be reset when the packet currently in the $j$\textsuperscript{th} position in the queue completes service, where $j=1$ corresponds to the packet in service. Note that since an update packet in the queue may not be of source $i$, $x_j(t)$ is in general not the age of the update itself. Also, in any discrete state $k$ we must track only the $k+1$ age processes $x_0(t), x_1(t),\ldots, x_k(t)$. The rest are irrelevant. We assume that in state $q(t) = k$, $x_j(t) = 0$ for $j > k$. The age state evolves according to 
\begin{align}
\dot{\xv}(t)=\bv_{k} \equiv \rvec{\onev[k+1]&\zerov[\n - k]}.
\end{align}
We use the shorthand notation $(\xv)_{j:l}$ and $(\vv_q)_{j:l}$, respectively, to denote the vectors $\rvec{x_j&\cdots&x_l}$ and $\rvec{v_{q,j} & \ldots & v_{q,l}}$. Table~\ref{tab:SSHResetTable} shows the transition reset maps for the transitions in the Markov chain in Figure~\ref{fig:CTMC}. We also show the age state  ${\xv_{q'}}$ and the vector ${\vv}_{q'}$ obtained on transitioning from state $q$ to $q'$.

\begin{table}
\caption{SSH transitions and reset maps for the chain in Figure~\ref{fig:CTMC}.}
\vspace{-8mm}
\begin{displaymath}\arraycolsep=1.0pt
{
\small
\begin{array}{ccccc}
q_l\to q'_l & \laml &  \xv\Amat_l & \Amat_l & \vv_{\ql}\Amat_l\\
\hline\\
1\to 0 &\mu	& \smvec{x_1&\mathbf{0}} &\resetMapArrDepart{0}&\smvec{v_{1,1}&\mathbf{0}}\\

0\to 1 &\lambdaOne	& \smvec{x_0&0&\mathbf{0}} &\resetMapArrSelf{1}&\smvec{v_{0,0}&0&\mathbf{0}}\\
0\to 1 &\lambdaTwo	& \smvec{x_0&x_{0}&\mathbf{0}} &\resetMapArrOther{1}&\smvec{v_{0,0}&v_{0,0}&\mathbf{0}}\\
2\to 1 &\mu	& \smvec{x_1 & x_2 &\mathbf{0}} &\resetMapArrDepart{1}&\smvec{v_{21} & v_{22} &\mathbf{0}}\\

\vdots&\vdots&\vdots&\vdots&\vdots\\

k-1\to k &\lambdaOne	& \smvec{(\xv)_{0:k-1}&0&\mathbf{0}} &\resetMapArrSelf{k}&\smvec{(\vv_{k-1})_{0:k-1}&0&\mathbf{0}}\\
k-1\to k &\lambdaTwo	& \smvec{(\xv)_{0:k-1}&x_{k-1}&\mathbf{0}} &\resetMapArrOther{k}&\smvec{(\vv_{k-1})_{0:k-1}&v_{k-1,k-1}&\mathbf{0}}\\
k+1\to k &\mu	& \smvec{(\xv)_{1:k+1}&\mathbf{0}} &\resetMapArrDepart{k}&\smvec{(\vv_{k+1})_{1:k+1}&\mathbf{0}}\\

\vdots&\vdots&\vdots&\vdots&\vdots\\

\n-1\to \n &\lambdaOne	& \smvec{(\xv)_{0:\n-1}&0&{0}} &\resetMapArrSelf{\n}&\smvec{(\vv_{\n-1})_{0:\n-1}&0&{0}}\\
\n-1\to \n &\lambdaTwo	& \smvec{(\xv)_{0:\n-1}&x_{\n-1}&{0}} &\resetMapArrOther{\n}&\smvec{(\vv_{\n-1})_{0:\n-1}&v_{\n-1,\n-1}&{0}}\\
\end{array}}
\end{displaymath}
\label{tab:SSHResetTable}
\vspace{-5mm}
\end{table}

Consider an arrival that transitions the system into state $0 < k < \n$ from $k-1$ as a result of an arrival from source $i$. The age processes $x_0(t), x_1(t), \ldots, x_{k-1}(t)$ stay unaffected by the transition and continue to evolve as in $k-1$. That is $x'_j = x_j$, $0 \le j \le k-1$. In state $k$ we must in addition track $x_k(t)$. We set $x_k(t) = 0$, since the age of this new update from source $i$ is $0$. Further, $\dot{x}_k(t) = 1$. The transition reset map is given by 
\begin{align}
\resetMapArrSelf{k} &= 
\left[
\begin{array}{c|c}
\begin{array}{c}
\underbracket{
	\begin{array}{ccccc}
	1&0&\cdots& 0 & 0\\
	0&1&\cdots& 0 & 0\\	
	\vdots & \vdots & \ddots & \vdots & \vdots\\
	0&0&\cdots& 1 & 0\\
	\end{array}
}_{k \times (k+1)} \\
{\mathbf{0}}
\end{array}&
\mathbf{0}
\end{array}
\right].
\end{align}
If the arrival is of another source, we must set $x_k(t)$ in a manner such that when this arrival completes service, it must not change the age of $i$ at the monitor. We set $x_k(t) = x_{k-1}(t)$. As before, $\dot{x}_k(t) = 1$. Note that in this case we are effectively setting $x_k(t)$ to the age of the freshest update of source $i$, if any, in the system. If there is no update of source $i$ in the system, we are effectively setting $x_k(t)$ to $x_0(t)$ as in this case all relevant age processes $x_0(t), x_1(t),\ldots, x_k(t)$ will be tracking the age at the monitor. The transition reset map is 
\begin{align}
\resetMapArrOther{k} &= 
\left[
\begin{array}{c|c}
\begin{array}{c}
\underbracket{
	\begin{array}{ccccc}
	1&0&\cdots& 0 & 0\\
	0&1&\cdots& 0 & 0\\	
	\vdots & \vdots & \ddots & \vdots & \vdots\\
	0&0&\cdots& 1 & 1\\
	\end{array}
}_{k \times (k+1)} \\
{\mathbf{0}}
\end{array}&
\mathbf{0}
\end{array}
\right].
\end{align}
Now consider a departure in state $k+1$. The SSH transitions to state $k$. The age process $x_1(t)$ in $k+1$ corresponds to the departure. On completion of service, this resets the process $x_0(t)$ at the monitor in state $k$. Once the packet departs, the one next in queue enters service. Thus the process $x_2(t)$ in state $k+1$ resets $x_1(t)$ in state $k$ and so on. The reset map is given by
\begin{align}
\resetMapArrDepart{k} &= 
\left[
\begin{array}{c|c}
\begin{array}{c}
\underbracket{
	\begin{array}{cccc}
	0&0&\cdots& 0\\
	1&0&\cdots& 0\\
	0&1&\cdots& 0\\
	\vdots & \vdots & \ddots & \vdots\\
	0&0&\cdots& 1
	\end{array}
}_{(k+2) \times (k+1)} \\
{\mathbf{0}}
\end{array}&
\mathbf{0}
\end{array}
\right].
\end{align}
We can now write the equations given by~\eqnref{AOI-SHS-v} in~\Thmref{AOI-SHS} for our blocking system.
\begin{subequations}
	\begin{align}
	\lambda \bar{\vv}_{0} &=\bv_{0}\bar{\pi}_{0}+ \mu \bar{\vv}_{1}\resetMapArrDepart{0}\eqnlabel{vq0},\\
	(\lambda + \mu) \bar{\vv}_{k} &=\bv_{k}\bar{\pi}_{k}+ \lambdaOne \bar{\vv}_{k-1}\resetMapArrSelf{k}\nonumber\\
	&\quad+ \lambdaTwo \bar{\vv}_{k-1}\resetMapArrOther{k} + \mu \bar{\vv}_{k+1} \resetMapArrDepart{k},\, 0 < k < \n,\eqnlabel{vqk}\\
	\mu \bar{\vv}_{\n} &=\bv_{\n}\bar{\pi}_{\n} + \lambdaOne \bar{\vv}_{\n-1}\resetMapArrSelf{\n} + \lambdaTwo \bar{\vv}_{\n-1}\resetMapArrOther{\n}\eqnlabel{vqm}.
	\end{align}
\end{subequations}
We extract equations corresponding to the discrete states $\bar q$ and the relevant age processes $x_j(t)$ in the states. We have $k+1$ relevant age processes in state $0 < k < \n$. For such a state $k$ and $x_j(t)$, $0 \le j \le k$, we have
\begin{subequations}
\begin{align}
(\lambda + \mu) \barv_{k,j} &= \bar{\pi}_k + \lambda \barv_{k-1,j}\nonumber\\
&\qquad+ \mu \barv_{k+1,j+1},\quad 0\le j\le k-1,\eqnlabel{vStateka}\\
(\lambda + \mu) \barv_{k,k} &= \bar{\pi}_k + \lambdaTwo \barv_{k-1,k-1} + \mu \barv_{k+1,k+1}.\eqnlabel{vStatekb}
\end{align}
\eqnlabel{vStatek}
\end{subequations}
In state $k=0$, no departures take place and the only relevant age process is $x_0(t)$. We have
\begin{align}
\lambda \barv_{0,0} = \bar{\pi}_0 + \mu \barv_{1,1}.
\eqnlabel{vState0}
\end{align}
Lastly, in state $k=M$, arrivals are inconsequential while a departure may take place. All age processes $x_0(t),\ldots,x_\n(t)$ are relevant. We have
\begin{subequations}
\begin{align}
\mu \barv_{\n,j} &= \bar{\pi}_\n + \lambda \barv_{\n-1,j},\, 0\le j\le n-1,\eqnlabel{vStateMa}\\
\mu \barv_{\n,\n} &= \bar{\pi}_\n + \lambdaTwo \barv_{\n-1,\n-1}.\eqnlabel{vStateMb}
\end{align}
\eqnlabel{vStateM}
\end{subequations}
The steady state probability $\bar{\pi}_k$ of $k$ updates in the blocking M/M/1 system can be obtained by solving~\eqnref{AOI-SHS-pi}-\eqnref{AOI-SHS-pi-sum}. The equations are
\begin{subequations}\eqnlabel{stat-probs}
\begin{align}
\lambda \bar{\pi}_0 &= \mu \bar{\pi}_1,\\
(\lambda + \mu) \bar{\pi}_k &= \lambda \bar{\pi}_{k-1} + \mu \bar{\pi}_{k+1},\, 0 < k < \n,\\
\mu \bar{\pi}_\n &= \lambda \bar{\pi}_{m-1}.
\end{align}
\end{subequations}
Combined with the normalization constraint $\sum_{q=0}^{\n} \bar{\pi}_q = 1$,  \eqnref{stat-probs} yields the M/M/1/$m$ stationary distribution 
\begin{align}
\bar{\pi}_k = \left(\frac{1-\rho}{1 - \rho^{\n+1}}\right) \rho^k,\quad 0 \le k \le \n.
\eqnlabel{steadyStateProb}
\end{align}

From~\Thmref{AOI-SHS} we know that the average age of source $i$ in the blocking system is 
\begin{align}
\age^\n_i = \sum_{\bar q=0}^\n \barv_{\bar q0}.
\eqnlabel{ageBlocking}
\end{align}
The average age $\age_i$ is obtained using Equation~\eqnref{age_as_limiting}. Details regarding calculation of $\age^\n_i$ and $\age_i$ are in the Appendix. With the definition
\begin{align}
\mathcal{E}_i \equiv \frac{1 + \rho - \sqrt{(1+\rho)^2 - 4\rho_{-i}}}{2\rho_{-i}},
\end{align}
we state our main result.
\begin{theorem}
Poisson sources $1,2,\ldots,N$ send their status updates to a monitor via an M/M/1 FCFS queue with service rate $\mu$. The sources offer loads $\rho_1,\ldots,\rho_N$. The average age of updates of source $i$ at the monitor is
\begin{align}
\age_i = \frac{1}{\mu}\left[\frac{1-\rho}{(\rho - \rho_{-i} \mathcal{E}_i) (1 - \rho \mathcal{E}_i)} + \frac{1}{1-\rho} + \frac{\rho_{-i}}{\rho_i}\right].\label{eqn:agetwouser}
\end{align}
\thmlabel{ageSourcei}
\end{theorem}
Note that the expression approaches the age of a single user M/M/1 queue as $\rho_{-i} \to 0$. This is easy to see on noting the fact that $\lim_{\rho_{-i} \to 0} \mathcal{E}_i = 1/(1+\rho)$.

\section{Evaluation}
\begin{figure}[t]
	\begin{center}
		\includegraphics[width=3.5in]{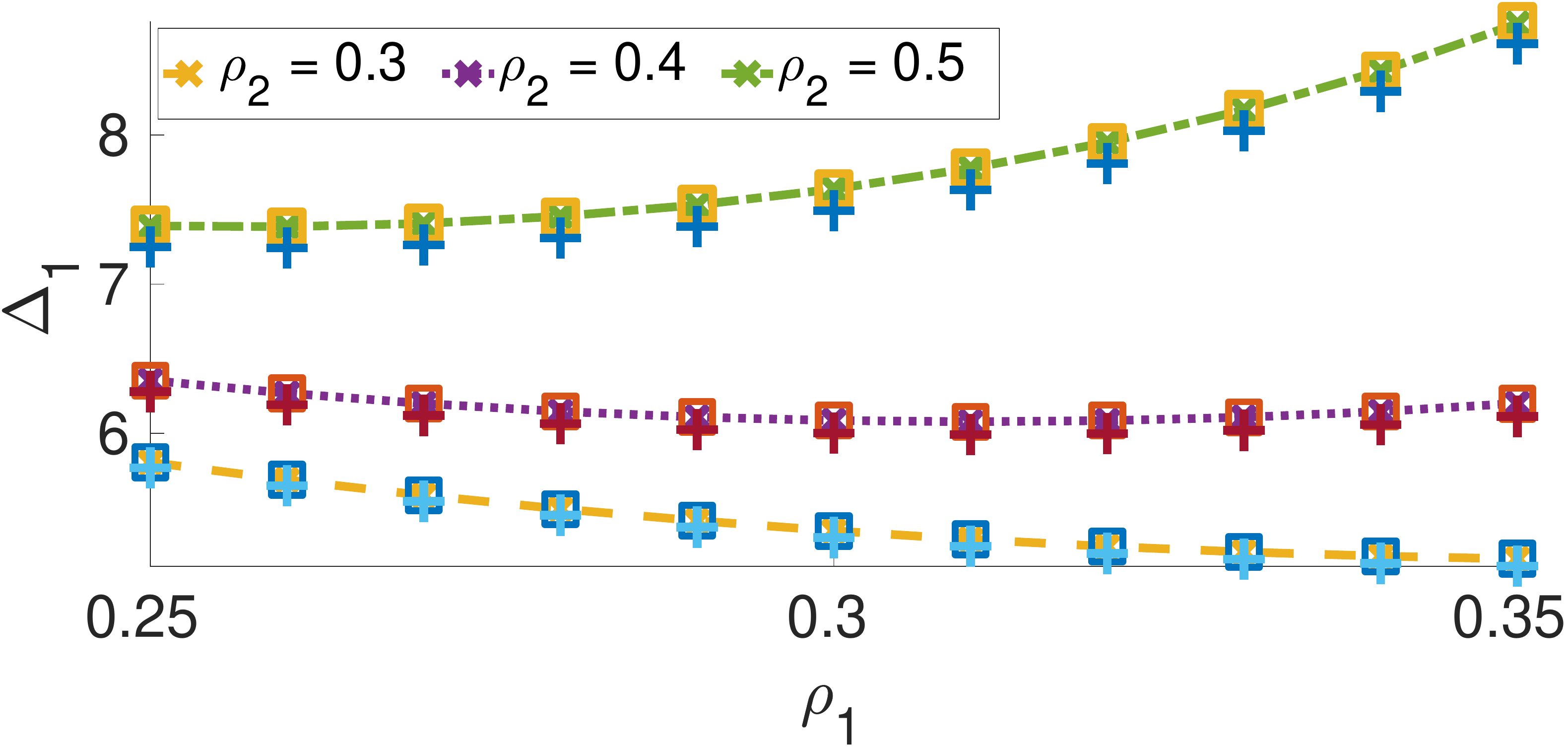}
		\caption{Age of source $1$ as a function of its utilization is shown for different values of utilization $\rho_2$ of the other source. For each $\rho_2$, the $\times$ markers correspond to the age given by~\eqnref{agetwouser}, the $\square$ markers show the age obtained using simulations, and the $+$ show the age obtained in the earlier work that analyzed multiple memoryless sources~\cite[Theorem $1$]{Yates-Kaul-IT2018}. While~\eqnref{agetwouser} matches well with simulations, the $+$ drift away as $\rho_2$ increases. We set $\mu=1$.}
		\label{fig:compareSimAndOld}	
	\end{center}
	\vspace{-5mm}		
\end{figure}

Figure~\ref{fig:compareSimAndOld} compares average age obtained using~\eqnref{agetwouser} with that estimated using simulation experiments. We plot the age $\age_1$ of source $1$ as a function of its offered load $\rho_1$ for various choices of the load $\rho_2$ offered by the other users. Each empirically estimated value of age was obtained by simulating on the order of $10^8$ transitions of the Markov chain. As is shown in the figure, empirical results match  our analysis. We also show the age that was obtained via analysis in our earlier work~\cite[Theorem $1$]{Yates-Kaul-IT2018}. We observe that the error in the age increases with the load of the other user. This error does disappear as $\rho_2 \to 0$. This is understandable as the error was introduced in accounting for the number of packets of other users in the system that are found by such an arrival of source $1$ that arrives after the previous arrival of the source has completed service.

Figure~\ref{fig:ageContours} shows average ages $\age_1$ and $\age_2$ of the sources as their relative loads are varied, for a fixed total load of $\rho$. The sum age is minimized when $\rho\approx 0.6$ and the sources share the load equally. At other relative loadings of the sources, the sum age may not be minimized at $\rho = 0.6$.

\section{Conclusion}
This work revisits average age analysis of status updates from multiple sources passing through the M/M/1 queue. SHS analysis of the M/M/1/$m$ blocking queue enables a relatively simple analysis and closed form result for the average age. 
\section*{Acknowledgement}
We thank the authors of \cite{moltafet2019age} for bringing the error in the M/M/1 analysis of \cite{Yates-Kaul-IT2018} to our attention.

\section*{Appendix: Proof of \Thmref{ageSourcei}}
\begin{proof}
We start by calculating the average $\age^{\n}_i$ of the blocking system as given in~\eqnref{ageBlocking}. We proceed by writing $\sum_{k=1}^\n \barv_{k,0}$ in terms of the probabilities $\bar{\pi}_k$, $0\le k \le \n$. That leaves us with the tedium of calculating $\barv_{0,0}$. Using~\eqnref{vStateka}, we can write
\begin{subequations}
	\begin{align}
	\lambda \barv_{0,0} &= \bar{\pi}_0 + \mu \barv_{1,1},\\
	\lambda \barv_{1,0} + \mu \barv_{1,0} - \lambda \barv_{0,0} &= \bar{\pi}_1 + \mu \barv_{21},
	\\
	&\ \, \vdots\nonumber\\
	\lambda \barv_{\n-1,0} + \mu \barv_{\n-1,0} - \lambda \barv_{\n-2,0} &= \bar{\pi}_{\n-2} + \mu \barv_{\n,1},\\
	\mu \barv_{\n,0} - \lambda \barv_{\n-1,0} &= \bar{\pi}_\n.
	\end{align}
\end{subequations}
Adding these equations we obtain
\begin{align}
\sum_{k=1}^\n \barv_{k,0} = \frac{1}{\mu}\sum_{k=0}^\n \bar{\pi}_k + \sum_{k=1}^{\n} \barv_{k,1}.
\eqnlabel{sumvk0}
\end{align}
Similarly, using~\eqnref{vStateka} to write equations corresponding to $\barv_{k,1}$, $1 \le k \le \n$, we obtain
\begin{align}
\sum_{k=1}^\n \barv_{k,1} = \frac{1}{\mu}\sum_{k=1}^\n \bar{\pi}_k + \frac{\lambdaTwo}{\mu} + \sum_{k=2}^{\n} \barv_{k,2}.
\end{align}
Proceeding similarly, we obtain expressions for $\sum_{k=2}^{\n} \barv_{k,2}, \sum_{k=3}^{\n} \barv_{k,3}, \ldots, \sum_{k=\n-1}^{\n} \barv_{k,\n-1}$. Further~\eqnref{vStateMb} gives us $\barv_{\n,\n}$. We can now rewrite~\eqnref{sumvk0} as
\begin{align}
\sum_{k=1}^\n \barv_{k,0} = \frac{1}{\mu}\sum_{k=0}^\n (k+1) \bar{\pi}_k + \frac{\lambdaTwo}{\mu} \sum_{k=0}^{\n-1} \barv_{k,k}.
\eqnlabel{sumvk0a}
\end{align}
Now consider equations~\eqnref{vStatekb} for $1 < k < m$,~\eqnref{vState0}, and~\eqnref{vStateMb}. Summing them gives
\begin{align}
\sum_{k=0}^{\n-1} \barv_{k,k} = \frac{1}{\lambdaOne}.
\eqnlabel{sumvkk}
\end{align}
Substituting in~\eqnref{sumvk0a} allows us to write $\sum_{k=1}^\n \barv_{k,0}$ in terms of the steady state probabilities given by~\eqnref{steadyStateProb}. To calculate $\barv_{0,0}$, we will express $\barv_{k,k}$ in terms of $\barv_{0,0}$ and then use~\eqnref{sumvkk}.

\begin{figure}[t]
	\begin{center}		
		\includegraphics[width=3.5in]{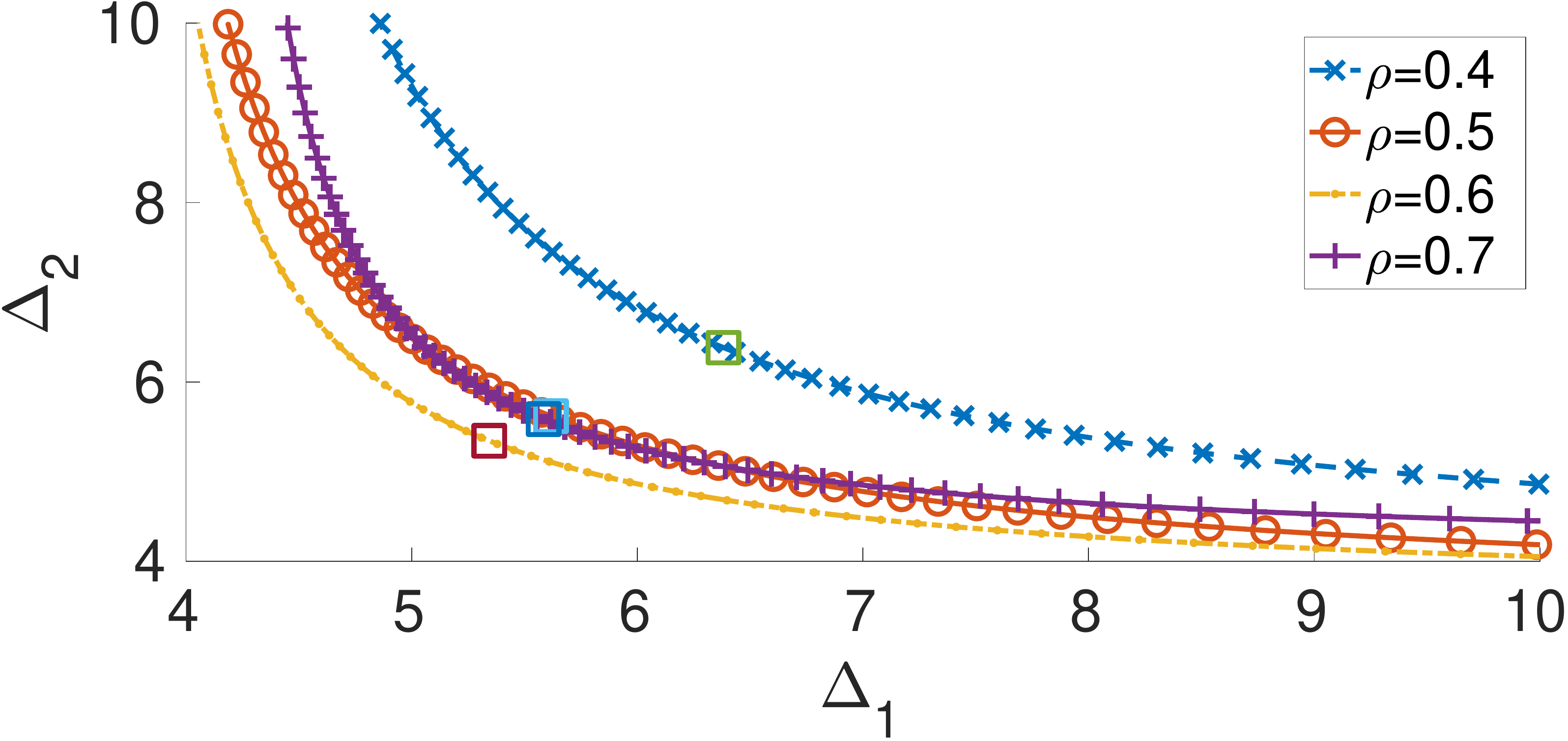}
		\caption{Age contours obtained by varying the share of two sources of a given total utilization $\rho$. The squares $\square$ correspond to the ages obtained when each source offers a load of $\rho/2$. Sum age is minimized when $\rho\approx 0.6$ for $\rho_1=\rho_2\approx 0.3$.}
		\label{fig:ageContours}	
	\end{center}
	\vspace{-5mm}		
\end{figure}

Rearranging terms in Equation~\eqnref{vStatekb} we get
\begin{align}
\barv_{k,k} \!=\! \left(\frac{\lambda + \mu}{\mu}\right) \barv_{k-1,k-1} \!-\! \left(\frac{\lambdaTwo}{\mu}\right)\barv_{k-2,k-2} \!-\! \frac{\bar{\pi}_{k-1}}{\mu}.
\end{align}
Repeated application of the equation to expand the terms $\barv_{k-1,k-1}$ and $\barv_{k-2,k-2}$ gives
\begin{align}
\barv_{k,k} = \left(\coeffOne_{1,k} \frac{\lambda}{\mu} - \coeffOne_{2,k} \frac{\lambdaTwo}{\mu}\right) \barv_{0,0} - \sum_{j=1}^{k} \coeffOne_{j,k}\frac{\bar{\pi}_{j-1}}{\mu}.
\eqnlabel{vkkInTermsofV00}
\end{align}
The coefficients $\coeffOne_{j,k}$ are obtained by solving the equations
\begin{align}
\coeffOne_{k-j,k} &= \frac{\lambda+\mu}{\mu} \coeffOne_{k-j+1,k} - \frac{\lambdaTwo}{\mu} \coeffOne_{k-j+2,k},
\end{align}
with $\coeffOne_{k,k} = 1$ and $0 \le j \le k$. We get
\begin{align}
\coeffOne_{j,k} &= \frac{(\beta - \rho_{-i}^{-1}(1+\rho)) \left((1+\rho)\rho_{-i}^{-1} + \beta\right)^{j-k}}{2\beta}\nonumber\\
  &\qquad+ \frac{(\beta + \rho_{-i}^{-1}(1+\rho))\left((1+\rho)\rho_{-i}^{-1} - \beta\right)^{j-k}}{2\beta},
\end{align}
where $\beta = \sqrt{(1+\rho)^2\rho_{-i}^{-2} - 4\rho_{-i}^{-1}}$.
Substituting $\coeffOne_{j,k}$ obtained above in~\eqnref{vkkInTermsofV00} and substituting the resulting $\barv_{k,k}$ in~\eqnref{sumvkk} gives us $\barv_{0,0}$. Together with~\eqnref{sumvk0a} and~\eqnref{sumvkk}, we can obtain the average age $\age^{\n}_i$ of the blocking system. Taking the limit as $m\to \infty$ gives us the average age $\age_i$ in Theorem~\thmref{ageSourcei}.	
\end{proof}

%
%

\begin{spacing}{0.983}
\bibliographystyle{IEEEtran}
\bibliography{AOI-2019-01}
\end{spacing}

\end{document}